\documentclass[sigconf]{acmart}

\usepackage{caption}
\usepackage{subcaption}
\usepackage{enumitem}
\usepackage{multirow}
\usepackage{bm}
\usepackage{hyperref}
\usepackage{makecell}

\AtBeginDocument{%
  \providecommand\BibTeX{{%
    \normalfont B\kern-0.5em{\scshape i\kern-0.25em b}\kern-0.8em\TeX}}}

\setcopyright{acmcopyright}
\copyrightyear{2024}
\acmYear{2024}
\acmDOI{XXXXXXX.XXXXXXX}

\acmConference[KDD '24]{Proceedings of the 30th ACM SIGKDD Conference On Knowledge Discovery and Data Mining}{August 25--29, 2024}{Barcelona, Spain}

\acmBooktitle{Proceedings of The 30th ACM SIGKDD Conference on Knowledge Discovery and Data Mining(KDD '24), August 25--29, 2024, Barcelona, Spain}

\begin{document}





\title[RIGL: A Unified Reciprocal Approach for Tracing the Independent and Group Learning Processes]{RIGL: A Unified Reciprocal Approach for Tracing the \\ Independent and Group Learning Processes }


\author{Xiaoshan Yu}
\affiliation{
  \institution{School of Artificial Intelligence, Anhui University}
  \city{Hefei}
  \state{Anhui}
  \country{China}
}
\authornote{Work was done at Career Science Lab, BOSS Zhipin supervised by Chuan Qin.}
\email{yxsleo@gmail.com}

\author{Chuan Qin}
\affiliation{
  \institution{Career Science Lab, BOSS Zhipin \\ PBC School of Finance, \\Tsinghua University}
  \city{Beijing}
  \country{China}
}
\authornotemark[2]
\email{chuanqin0426@gmail.com}

\author{Dazhong Shen}
\affiliation{
  \institution{Shanghai Artificial Intelligence Laboratory}
  \city{Shanghai}
  \country{China}
}
\email{dazh.shen@gmail.com}

\author{Shangshang Yang}
\affiliation{
  \institution{School of Artificial Intelligence, Anhui University}
  \city{Hefei}
  \state{Anhui}
  \country{China}
}
\email{yangshang0308@gmail.com}

\author{Haiping Ma}
\affiliation{
  \institution{Information Materials and Intelligent Sensing Laboratory of Anhui Province, Institutes of Physical Science and Information Technology, Anhui University}
  \city{Hefei}
  \state{Anhui}
  \country{China}
}
\authornotemark[2]
\email{hpma@ahu.edu.cn}


\author{Hengshu Zhu}
\affiliation{
  \institution{Career Science Lab, BOSS Zhipin}
  \city{Beijing}
  \country{China}
}
\authornote{Corresponding authors.}
\email{zhuhengshu@gmail.com}

\author{Xingyi Zhang}
\affiliation{
  \institution{School of Computer Science and Technology, Anhui University}
  \city{Hefei}
  \state{Anhui}
  \country{China}
}
\email{xyzhanghust@gmail.com}

\begin{abstract}

In the realm of education, both independent learning and group learning are esteemed as the most classic paradigms. The former allows learners to self-direct their studies, while the latter is typically characterized by teacher-directed scenarios. Recent studies in the field of intelligent education have leveraged deep temporal models to trace the learning process, capturing the dynamics of students' knowledge states, and have achieved remarkable performance. However, existing approaches have primarily focused on modeling the independent learning process, with the group learning paradigm receiving less attention. Moreover, the reciprocal effect between the two learning processes, especially their combined potential to foster holistic student development, remains inadequately explored.
To this end, in this paper, we propose \textbf{RIGL}, a unified \underline{\textbf{R}}eciprocal model to trace knowledge states at both the individual and group levels, drawing from the \underline{\textbf{I}}ndependent and \underline{\textbf{G}}roup \underline{\textbf{L}}earning processes.
Specifically, we first introduce a time frame-aware reciprocal embedding module to concurrently model both student and group response interactions across various time frames. Subsequently, we employ reciprocal enhanced learning modeling to fully exploit the comprehensive and complementary information between the two behaviors. Furthermore, we design a relation-guided temporal attentive network, comprised of dynamic graph modeling coupled with a temporal self-attention mechanism. It is used to delve into the dynamic influence of individual and group interactions throughout the learning processes, which is crafted to explore the dynamic intricacies of both individual and group interactions during the learning sequences. Conclusively, we introduce a bias-aware contrastive learning module to bolster the stability of the model's training. Extensive experiments on four real-world educational datasets clearly demonstrate the effectiveness of the proposed RIGL model. Our codes are available at \url{https://github.com/LabyrinthineLeo/RIGL}.

\end{abstract}

\renewcommand{\shortauthors}{Xiaoshan Yu et al.}






\maketitle



\section{Introduction}





In the domain of education, both independent learning~\cite{IndLearning} and group learning~\cite{GroupLearning,RDGT} are regarded as the most classic learning paradigms. The former allows learners to self-direct their studies, whereas the latter is typically characterized by scenarios that are guided and structured by teachers. It is widely acknowledged that exclusive reliance on a singular learning modality is insufficient to promote continuous long-term development in students~\cite{Group_Ind_1, Group_Ind_2}.


\begin{figure}[!t] 
    \centering 
    \includegraphics[width=0.95\linewidth]{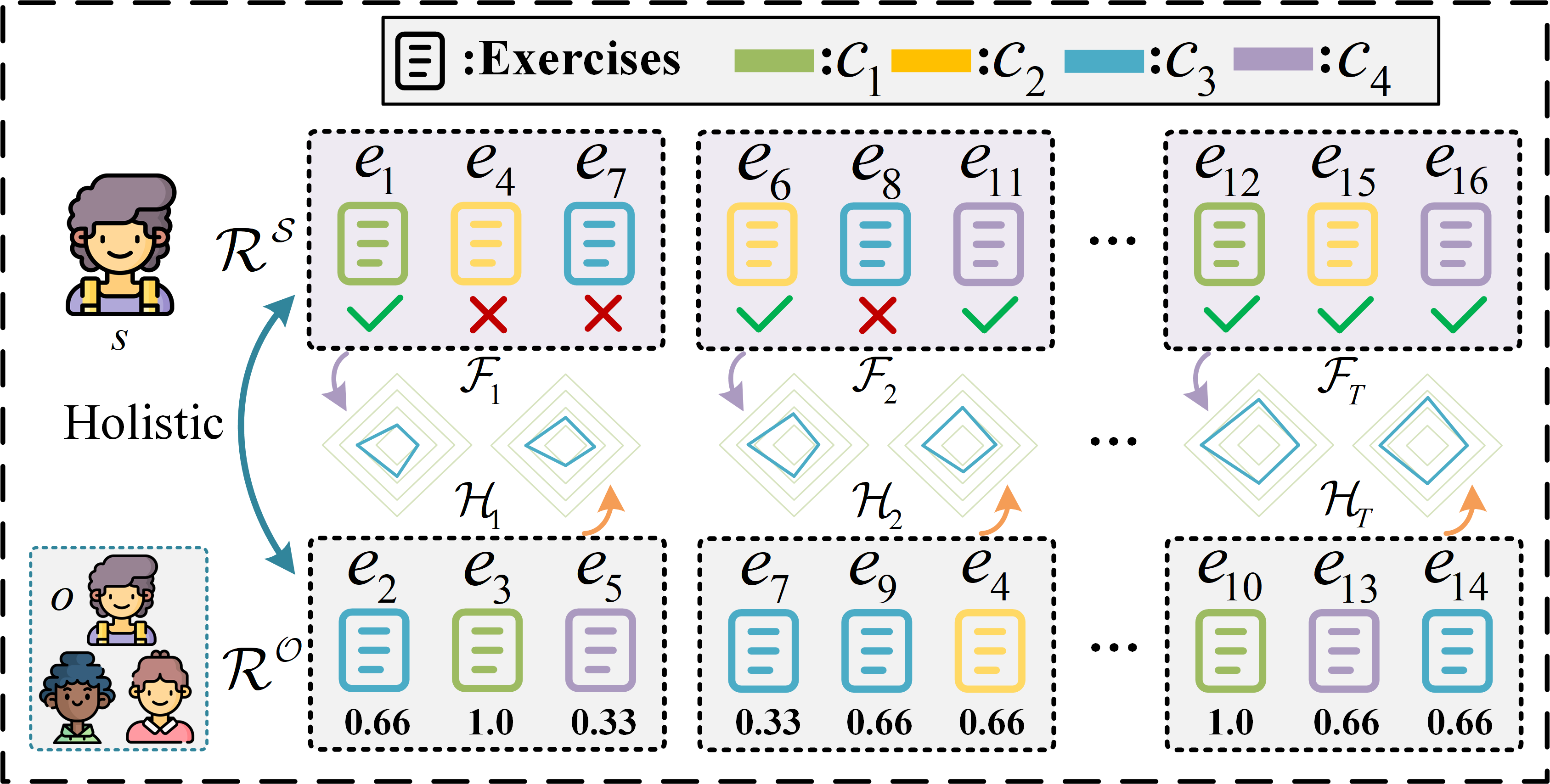} 
    \vspace{-2mm}
    \caption{An illustrative example of the holistic knowledge tracing (HKT) task. The top and bottom halves indicate the individual and group learning processes, respectively, which are organized in time frames, and the radar chart in the middle represents the knowledge proficiency levels of both. }
    \label{fig.gcd} 
    \vspace{-6mm}
\end{figure}


In recent years, with the advancement of artificial intelligence technology, the field of intelligent education has yielded notable modeling paradigms conducive to understanding student learning behaviors~\cite{AIED,liu2023homogeneous,ma2022prerequisite}. Among these, a foundational and potent paradigm is knowledge tracing, which aims at dynamically monitoring the evolving knowledge states of learners and predicting their future performance by modeling the exercise-solving sequences~\cite{KT}. However, most existing methods~\cite{DKT,SAKT,AKT,ENAS-KT,HD-KT} mainly focus on modeling independent learning behaviors, with the group learning paradigm receiving less attention. Furthermore, the reciprocal effect between independent and group learning, particularly their combined potential to significantly drive holistic student development, has yet to be thoroughly explored and investigated~\cite{Group_Ind_2,yang2023cognitive,yang2023designing,yang2023evolutionary}.

To this end, in this paper, we introduce a new task called holistic knowledge tracing (HKT), which refers to tracing knowledge states at both the individual and group levels simultaneously, drawing from independent and group learning processes. As shown in Figure~1, 
$\mathcal{R}^{\mathcal{S}}$ and  $\mathcal{R}^{\mathcal{O}}$ indicate the interaction sequences for a student and a group respectively,
which are organized in time frames. Each response interaction of a student and a group under each time frame is represented by triples $(e^t_i,c^t_i,r^t_i)$ and $(e^t_i,c^t_i,y^t_i)$, respectively, where $e^t_i$ and $c^t_i$ denote the exercise and the knowledge concept, and $r^t_i \!\in \!\{0,\!1\}$ as well as $y^t_i\!\in\![0,\!1]$ indicates the student's response and the group's answer accuracy rate. The goal of HKT is to model both learning~processes~holistically.

However, HKT is not a trivial task and encompasses the following technical challenges. Firstly, within the real-world educational environment, there exists a notable absence of interactive behaviors among students spanning various learning scenarios, particularly within group learning processes. This absence significantly amplifies the challenge of tracing knowledge states at both individual and group levels. For instance, missing data resulting from the absence of a student from a class test can cause bias in the overall assessment. Secondly, in contrast to traditional knowledge tracing that only focuses on individuals, the HKT task requires simultaneously and effectively modeling learning processes while exploring the dynamic interactions between individuals and groups during the learning journey, which is quite confronting.

To address these challenges, in this paper, we propose a unified \underline{\textbf{R}}eciprocal approach to trace the \underline{\textbf{I}}ndependent and \underline{\textbf{G}}roup \underline{\textbf{L}}earning processes (\textbf{RIGL}), aimed at delivering a productive dynamic assessment for both students and groups. Specifically, we first design a time frame-aware reciprocal embedding module to simultaneously model students' and groups' response interactions over time frames and then used the reciprocal enhanced learning modeling to fully exploit the comprehensive and complementary information between the two behaviors. 
Subsequently, we propose a relation-guided temporal attentive network comprised of dynamic graph modeling and a temporal self-attentive mechanism for exploring the dynamic complexity of individual and group interactions during the learning processes. In particular, the relation-guided dynamic graph is constructed by mining potential associations between students and groups. Finally, a bias-aware contrastive learning module is introduced to ensure the stability of training. Extensive experiments on four real-world educational datasets clearly demonstrate the effectiveness of the proposed RIGL model in the HKT task.

\vspace{-3mm}
\section{Related Work}

\subsection{Knowledge Tracing}

Knowledge tracing (KT)~\cite{KT} aims to monitor the changing knowledge states of learners by modeling their exercise-solving sequences as a sequence prediction task, which has been recognized as an immensely crucial research task in the field of intelligent education. Over the past decades, numerous effective KT models have been proposed. Among these, the traditional KT approaches play an essential role, which usually utilizes probabilistic models~\cite{KT,ma2022reconciling,zhang2023relicd} or logistic functions~\cite{PFA,BKT_survey,yang2024endowing} to model the knowledge states of students. 
In recent years, the rapid advancements of deep learning have propelled neural network (NN)-based KT approaches into the dominant paradigm~\cite{DKT,DKVMN,SAKT,AKT,simpleKT}. 
These approaches leverage the power of neural networks to dynamically mine the knowledge acquisition process of students by solving the sequence prediction task, by which performance improvement and personalized educational experiences are achieved. For instance, DKT~\cite{DKT} utilizes a recurrent neural network (RNN) to model the student's exercising sequence and represent student cognitive proficiency with the hidden states. 
In particular, DKVMN~\cite{DKVMN} introduces the memory-augmented neural network into KT, which defines two matrices called \textit{key} and \textit{value} to store and update student's knowledge mastery, respectively. Furthermore, SAKT~\cite{SAKT} exploits the transformer architecture~\cite{Transformer} to explore long-term relations of interaction behaviors in students' learning history for the first time.
Despite the success of these approaches, they primarily focus on individual assessment and thus leave a gap in the availability of a holistic knowledge tracing framework that simultaneously models individual and group learning behaviors.

\vspace{-3mm}
\subsection{Dynamic Graph Representation Learning}

Dynamic graph representation learning~\cite{DynGraph_survey3,zha2023career} is a rapidly evolving field that focuses on effectively capturing temporal dependencies and changing patterns within dynamic graph-structured data. In recent years, many approaches have been proposed to effectively model and learn the structural information and representation of dynamic graphs in various research problems, such as link prediction~\cite{DynGraph_LinkPred,qin2023comprehensive,qin2023automatic,sun2021market}, knowledge retrieval~\cite{jiang2024enhancing,jiang2024towards}, and career development~\cite{fang2023recruitpro,qin2019duerquiz,zheng2024bilateral,yao2023resuformer}. One classical category of approaches is to conceptualize dynamic graphs by dividing them into multiple graph snapshots with discrete time characteristics~\cite{DynGraph_cat1_dysat,DynGraph_cat1_evogcn,chen2024collaboration}. For example, DySAT~\cite{DynGraph_cat1_dysat} employs a dual-dimension self-attention mechanism, combining structural attention for local node features in graph snapshots with temporal attention to track graph evolution, enhanced by multiple attention heads for diverse graph structure analysis. EvolveGCN~\cite{DynGraph_cat1_evogcn} introduces a dynamic adaptation of the graph convolutional network (GCN) model across time, using an RNN to evolve its parameters and capture the dynamics of graph sequences, with two different architectures for parameter evolution. 

Another avenue of exploration regards time as a continuous feature, treating the dynamic graph as a stream of timestamped events to derive node representations~\cite{DynGraph_cat2_dyrep,DynGraph_cat2_temp,DynGraph_LinkPred}. For instance, DyRep~\cite{DynGraph_cat2_dyrep} is a dynamic graph framework conceptualizing representation learning as a latent mediation process bridging two observed processes namely–dynamics of the network and dynamics on the network, which leverages a two-time scale deep temporal point process and a temporal-attentive network to intertwine network topology and node activity dynamics. Moreover, dynamic graph learning are also applied in intelligent tutoring systems (ITS), e.g., TEGNN~\cite{TEGNN} presents a method that combines a heterogeneous evolution network with a temporal extension graph neural network to dynamically model entities and relations~\cite{fang2023recruitpro} in the intelligent tutoring system. Although these strategies behave well in many tasks, how to introduce this idea into holistic knowledge tracing, where the knowledge states of individuals and groups vary dynamically over time and the associations and influences between them are difficult to construct and capture directly, has not been explored.

\begin{figure*}[!t] 
	\centering 
        \vspace{-2mm}
	\includegraphics[scale=0.87]{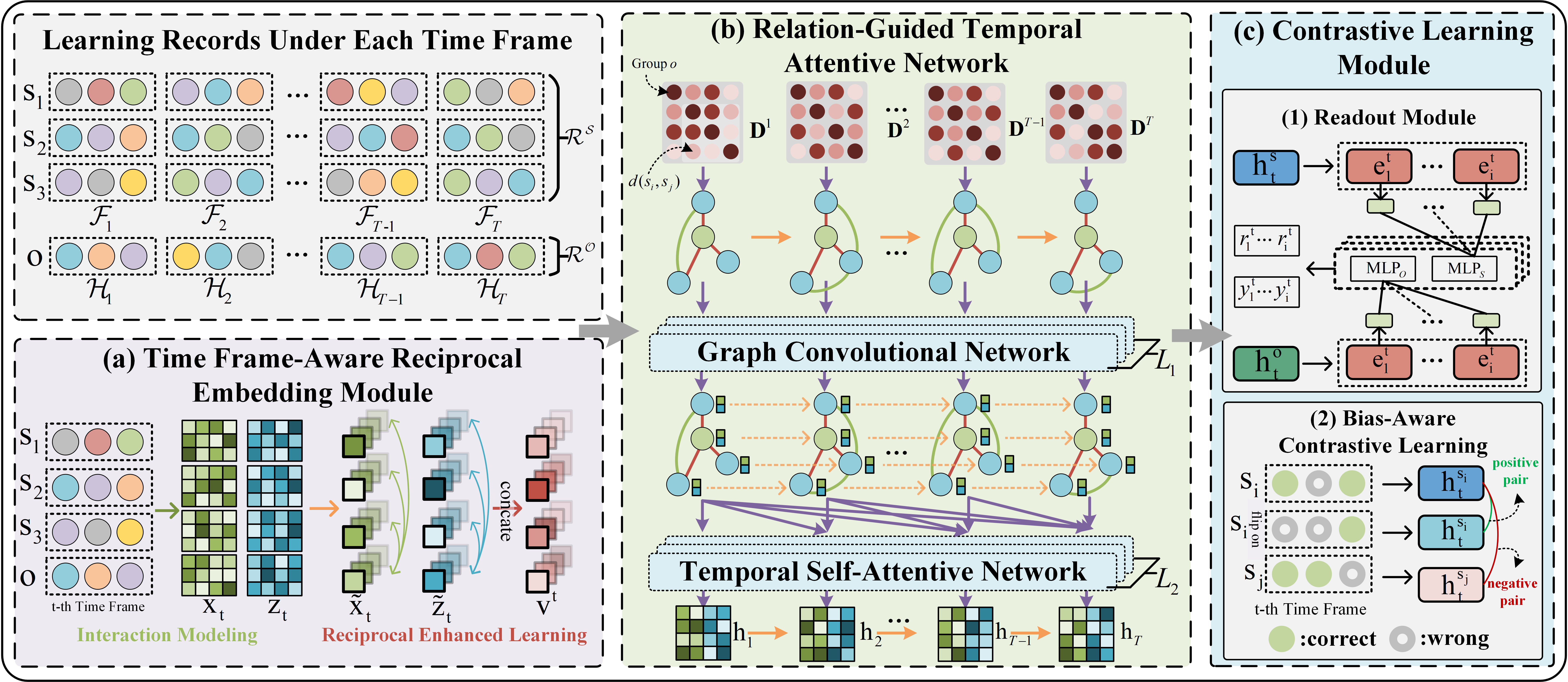} 
    \vspace{-2mm}
    \caption{The overview architecture of our proposed RIGL model. (a) The time frame-aware reciprocal embedding module includes both individual and group interaction modeling as well as reciprocal enhanced learning. (b) The relation-guided temporal attentive network models the complex learning processes with dynamic changing knowledge, including the relation-guided dynamic graph modeling and a temporal self-attentive network. (c) The contrastive learning module generates the augmented student interactions by randomly flipping responses considering the learning bias during exercise solving, such as carelessness or guessing, and promotes the training stability through this bias-aware contrastive learning. Best viewed in color.}
        \label{fig.framework} 
        \vspace{-5mm}
\end{figure*}

\vspace{-3mm}
\section{Problem Formulation}
In this section, we formally define the holistic knowledge tracing (HKT) problem. Let $\mathcal{O} \!=\! \{o_1,\ldots,o_I\}$ be the set of $I$ groups, $\mathcal{S} = \{s_1,\ldots,s_N\}$ be the set of $N$ students, $\mathcal{E} \!=\! \{e_1,\ldots,e_M\}$ be the set of $M$ exercises, and $\mathcal{C} \!=\! \{c_1,\ldots,c_K\}$ be the set of $K$ knowledge concepts. The relationship between exercises and concepts is denoted by $Q$-matrix $Q = \{q_{ij}\}^{M\times K}$, where $q_{ij} = 1$ if exercise $e_i$ requires concept $c_j$ and $0$ otherwise. Each group consists of a certain number of students, e.g., the $i$-th group $o_i = \{s^i_{1},\ldots,s^i_{|o_i|}\}$, where $s^i_{*} \in \mathcal{S}$ and $|o_i|$ is the size~of~group~$o_i$.  

Generally, students perform personalized learning activities under time frames, i.e., answering a certain number of exercises at specific time intervals. We denote the whole interaction sequence of a student with $T$ time frames as $\mathcal{R}^{\mathcal{S}} = \{\mathcal{F}_1,\ldots, \mathcal{F}_T\}$, where $\mathcal{F}_t = \{(e^t_1,c^t_1,r^t_1), \ldots,(e^t_{|\mathcal{F}_t|},c^t_{|\mathcal{F}_t|},r^t_{|\mathcal{F}_t|}) \}$ stands for the interaction sequence under $t$-th time frame; the triple $(e^t_i,c^t_i,r^t_i)$ refers $i$-th exercising record; $e^t_i \in \mathcal{E}$ is the exercise; $c^t_i \in \mathcal{C}$ is the concept associated with the exercise $e^t_i$, which is obtained from the $Q$; and $r^t_i \in \{0,1\}$ is the response score. Meanwhile, students would engage in collective learning behaviors under time frames, where all students in the group completed the same batch of exercises. We denote the whole group interaction sequence for a group with $T$ time frames as $\mathcal{R}^{\mathcal{O}} = \{\mathcal{H}_1, \ldots, \mathcal{H}_T\}$, where $\mathcal{H}_t = \{(e^t_1,c^t_1,y^t_1), \ldots,(e^t_{|\mathcal{H}_t|},c^t_{|\mathcal{H}_t|},y^t_{|\mathcal{H}_t|}) \}$ is the group-exercise interaction sequence under $t$-th time frame; the triple $(e^t_i,c^t_i,y^t_i)$ denotes $i$-th interaction log; $e^t_i \in \mathcal{E}$, $c^t_i \in \mathcal{C}$ and $y^t_i \in [0,1]$ is the correct rate that the group got.

\noindent \textbf{Problem Definition.} \textit{Given student's interaction sequence $\mathcal{R}^{\mathcal{S}} = \{\mathcal{F}_1,\ldots,\mathcal{F}_T\}$, where $\mathcal{F}_t = \{(e^t_1,c^t_1,r^t_1),\ldots,(e^t_{|\mathcal{F}_t|},c^t_{|\mathcal{F}_t|},r^t_{|\mathcal{F}_t|}) \}$ and group's interaction sequence $\mathcal{R}^{\mathcal{O}} = \{\mathcal{H}_1,\ldots, \mathcal{H}_T\}$, where $\mathcal{H}_t = \{(e^t_1,c^t_1,y^t_1),\ldots,(e^t_{|\mathcal{H}_t|},c^t_{|\mathcal{H}_t|},y^t_{|\mathcal{H}_t|}) \}$, the goal of holistic knowledge tracing is twofold: (1) simultaneously diagnosing the knowledge state of each student within a group and the group-level proficiency of the corresponding group from time frame $t_1$ to $t_T$; and (2) simultaneously predicting students' performance scores as well as the group's correct rate on specific exercises at time frame~$t_{T+1}$.} Notably, unlike traditional KT task, the interaction elements (e.g., $(e^t_i,c^t_i,y^t_i) \in \mathcal{F}_t$) under each current time frame in the HKT task are not used for prediction to avoid information leakage.



\vspace{-3mm}
\section{Methodology}
In this section, we present the RIGL model in detail. As illustrated in Figure 2, the architecture of RIGL mainly consists of three components, which are a time frame-aware reciprocal embedding module, a relation-guided temporal attentive network, and a contrastive learning module. 

\vspace{-4mm}
\subsection{Time Frame-Aware Reciprocal Embedding Module}

Effectively representing student interaction and modeling knowledge acquisition during the learning process has always been very critical in traditional knowledge tracing task~\cite{LPKT}. Similarly, to effectively model the learning process of students and groups under each time frame in the HKT task, we carefully design three sub-modules: the individual interaction modeling, the group interaction modeling, and the reciprocal enhanced learning module, which are detailed in the following.

\subsubsection{Individual Interaction Modeling}
As mentioned earlier, each student typically solves multiple exercises under each time frame $\mathcal{F}_t$. Each interaction behavior $(e^t_i,c^t_i,r^t_i)$ contains the interactive exercises, the involved knowledge concept, and the corresponding response, which are often rich in information~\cite{AKT}. We first encode the question and response 
 for each student interaction $i$, as follows:
\begin{equation}
    \small
    \label{Encoder_stu}
    \begin{split}
        \textbf{x}_i^t = \textbf{[}\textbf{e}_i^t \oplus \textbf{c}_i^t\textbf{]} \; \textbf{W}_1 + \textbf{b}_1; \; \textbf{z}_i^t = \textbf{r}_i^t,
    \end{split}
\end{equation}
where $\textbf{e}_i^t \in \mathbb{R}^{d_e}$, $\textbf{c}_i^t \in \mathbb{R}^{d_c}$, and $\textbf{r}_i^t \in \mathbb{R}^{d_r}$ denote the latent representations of $e_i^t$, $c_i^t$, and $r_i^t$ respectively, $\textbf{W}_1 \in \mathbb{R}^{d_e \times d}$ and $\textbf{b}_1 \in \mathbb{R}^{d}$ are the trainable parameters, and $\oplus$ refers to the element-wise addition operator. Notably, $d_e$, $d_c$, and $d_r$ are the dimensions of the embeddings of exercise, concept, and response respectively, and here $d_e$ equals to $d_c$.

After the student interaction encoding, we obtain a set of exercise encoding $\textbf{x}^t = \{\textbf{x}_1^t, \textbf{x}_2^t, \ldots, \textbf{x}_{|\mathcal{F}_t|}^t\}$ and a set of response encoding $\textbf{z}^t = \{\textbf{z}_1^t, \textbf{z}_2^t, \ldots,\textbf{z}_{|\mathcal{F}_t|}^t\}$ under the time frame $\mathcal{F}_t$. Considering that a student's capability is usually stable over a short period of time~\cite{QuizKT}, we utilize all of the student's exercising behaviors under a time frame to comprehensively model her knowledge acquisition. Specifically, we leverage the average pooling operation as a knowledge aggregator~\cite{QuizKT} to fuse all interactions thereby perceiving knowledge gain during student learning as below:
\begin{equation}
    \small
    \label{Encoder_Agg_stu}
    \begin{split}
        \textbf{x}_t^s = \frac{1}{|\mathcal{F}_t|}\sum\limits_{i=1}^{|\mathcal{F}_t|} \textbf{x}_i^t; \; \textbf{z}_t^s = \frac{1}{|\mathcal{F}_t|}\sum\limits_{i=1}^{|\mathcal{F}_t|} \textbf{z}_i^t,
    \end{split}
\end{equation}
where $\textbf{x}_t^s \in \mathbb{R}^{d}$ and $\textbf{z}_t^s \in \mathbb{R}^{d}$ denote the knowledge representation and the performance representation of student $s$ under $t$-th time frame, respectively.



\subsubsection{Group Interaction Modeling}
Similar to the exercise-solving process of students, there will be multiple group-exercise interaction records under one time window. Given any one interaction $(e_i^t, c_i^t,y_i^t)$ under the $t$-th time frame $\mathcal{H}_t$ of group $o$, it consists of the question, the concept, and the percentage of correct responses. We first encode the exercise traits and the collective response information:
\begin{equation}
    \small
    \label{Encoder_grp}
    \begin{split}
        \hat{\textbf{x}}_i^t = \textbf{[}\textbf{e}_i^t \oplus \textbf{c}_i^t\textbf{]} \; \textbf{W}_2 + \textbf{b}_2; \; \hat{\textbf{z}}_i^t = y_i^t \textbf{h}_1 + b_1,
    \end{split}
\end{equation}
where $\textbf{e}_i^t \in \mathbb{R}^{d_e}$ and $\textbf{c}_i^t \in \mathbb{R}^{d_c}$ refer to the embeddings of $e_i^t$ and $c_i^t$ respectively, $\textbf{W}_2 \in \mathbb{R}^{d_e \times d}$, $\textbf{b}_2 \in \mathbb{R}^{d}$, $\textbf{h}_1 \in \mathbb{R}$, and $b_1 \in \mathbb{R}$ are the trainable parameters. 
Note that we set the new parameters $\textbf{W}_2$ and $\textbf{b}_2$ different from Eq.~\ref{Encoder_stu} to model the question embeddings under the perspective of the student group. 
After acquiring the group interaction encoding sets $\hat{\textbf{x}}^t = \{\hat{\textbf{x}}_1^t, \hat{\textbf{x}}_2^t, \ldots, \hat{\textbf{x}}_{|\mathcal{H}_t|}^t\}$ and $\hat{\textbf{z}}^t = \{\hat{\textbf{z}}_1^t, \hat{\textbf{z}}_2^t, \ldots,\hat{\textbf{z}}_{|\mathcal{H}_t|}^t\}$, we further model the learning evolution of group level within the time frame, similar to Eq.(2):
\begin{equation}
    \small
    \label{Encoder_Agg_grp}
    \begin{split}
        \textbf{x}_t^o = \frac{1}{|\mathcal{H}_t|}\sum\limits_{i=1}^{|\mathcal{H}_t|} \hat{\textbf{x}}_i^t; \; \textbf{z}_t^o = \frac{1}{|\mathcal{H}_t|}\sum\limits_{i=1}^{|\mathcal{H}_t|} \hat{\textbf{z}}_i^t.
    \end{split}
\end{equation}

\subsubsection{Reciprocal Enhanced Learning}
Individual and group learning are always interrelated and complementary in organizations (e.g., classes, teams~\cite{lin2017collaborative}), where the collective activities in which students participate contribute to the complementation of the students' knowledge proficiency, as well as the individualized learning history of the students promotes the perception of the group-level ability~\cite{Group_Ind_1,DGCD}. Inspired by this, we propose a reciprocal enhanced learning module for tracing individual and group interaction simultaneously (as shown in Figure~2(a)). Specifically, we first utilize group interaction information to enrich individual learning features:
\begin{equation}
    \small
    \label{Encoder_Reci_stu}
    \begin{split}
        \widetilde{\textbf{x}}_t^s = \textbf{x}_t^s \oplus \textbf{x}_t^o; \; \widetilde{\textbf{z}}_t^s = \textbf{z}_t^s \oplus \textbf{z}_t^o,
    \end{split}
\end{equation}
where $\widetilde{\textbf{x}}_t^s, \widetilde{\textbf{z}}_t^s \in \mathbb{R}^d$ stand for the enhanced student interaction encoding. As well, we utilize personalized learning behaviors of students within the same group to enhance the modeling of group-level learning interaction. In particular, to address the challenge of individual absence in collective interaction, we elaborate an absence-perceived attention aggregation module:
\begin{equation}
    \small
    \label{Encoder_Reci_grp}
    \begin{split}
        \widetilde{\textbf{x}}_t^o = \textbf{x}_t^o \oplus \sum\limits_{i=1}^{|o|}\lambda_i \textbf{x}_t^{s_i}; \; \widetilde{\textbf{z}}_t^o = \textbf{z}_t^o \oplus \sum\limits_{i=1}^{|o|}\lambda_i \textbf{z}_t^{s_i},
    \end{split}
\end{equation}
where $\textbf{x}_t^{s_i}, \textbf{z}_t^{s_i} \in \mathbb{R}^d$ are the interaction features of $i$-th student $s_i$ within group $o$, and $\lambda_i$ denotes the absence-perceived contribution weight, which is computed by, 
\begin{equation}
    \small
    \label{Encoder_Reci_lambda}
    \begin{split}
        \hat{\lambda}_i &= ReLU(\textbf{[}\textbf{x}_t^{s_i};\textbf{z}_t^{s_i}\textbf{]} \textbf{W}^k + \textbf{[}\textbf{x}_t^o;\textbf{z}_t^o\textbf{]} \textbf{W}^q)\textbf{h}_2, \\
        \lambda_i &= softmax(\hat{\lambda}_i) = \frac{exp(\hat{\lambda}_i)}{\sum_{j=1}^{|o|}exp(\hat{\lambda}_j)}.
    \end{split}
\end{equation}
where $\textbf{W}^k,\textbf{W}^q \in \mathbb{R}^{2d\times d}$ are the key and query matrices of the attention layer, $\textbf{h}_2 \in \mathbb{R}^d$ is the weight vector for projecting attention scores, and $softmax(\cdot)$ and $\textbf{[} \bm{\cdot} \; ; \; \bm{\cdot} \textbf{]}$ denote the softmax function and the concatenation operation.

\subsection{Relation-Guided Temporal Attentive Network}

Considering the dynamic complexity of the learning process of students and groups and the information interaction between them, in this section, we propose a relation-guided temporal attentive network to model this complex learning process with dynamic changing knowledge, which consists of the relation-guided dynamic graph modeling and a temporal self-attentive network.

\subsubsection{Relation-Guided Dynamic Graph Modeling}
In this section, we first describe how to construct the relation-guided dynamic graph and then design a dynamic GCN module to enhance the relation modeling. Referring to previous work~\cite{DynGraph_cat1_dysat}, we consider the global dynamic graph as a series of static graph snapshots, i.e., $\mathcal{G} = \{\mathcal{G}_1, \mathcal{G}_2, \ldots, \mathcal{G}_T\}$, where $T$ is the number of time frames. 
For time frame $t \in T$, with respect to the personalized learning behaviors of the students in the group $o \in \mathcal{O}$ and the interactive response behaviors of the group, we construct the corresponding group-individual graph $\mathcal{G}_t^o = (V, E^t)$ with a node set $V = \{o, s_1, \ldots, s_{|o|}\}$ and a edge set $E^t = (E_{o \leftrightarrow s}, E_{s \leftrightarrow s}^t)$, where $E_{o \leftrightarrow s}$ denotes the set of edges connecting the group node $o$ and all other student nodes, and $E_{s \leftrightarrow s}^t$ denotes the set of connecting edges between students under time stage $t$, which is dynamically changing.




Specifically, a relation-guided approach is proposed to construct the changing edges between students. We first incorporate the above interaction encoding as node feature:
\begin{equation}
    \small
    \label{Encoder_nodes}
    \begin{split}
        \textbf{v}_0^t = \textbf{[}\widetilde{\textbf{x}}_t^o;\widetilde{\textbf{z}}_t^o\textbf{]};  \textbf{v}_i^t = \textbf{[}\widetilde{\textbf{x}}_t^{s_i};\widetilde{\textbf{z}}_t^{s_i}\textbf{]}, i \in \{1,2,\ldots,|o|\},
    \end{split}
\end{equation}
where $\textbf{v}_0^t, \textbf{v}_i^t \in \mathbb{R}^{2d}$ denote the feature vector of group node and student nodes respectively. Then, we obtain the node feature matrix $\textbf{V}^t \in \mathbb{R}^{(|o|+1)\times 2d}$. For any two student nodes $\mathit{v}_i^t$ and $\mathit{v}_j^t$ in the time frame $t$, where $i, j \in \{1,2,\ldots,|o|\}$, we acquire the relation distance by calculate the exercise interaction similarity between them and then obtain the relation matrix as follows: 
\begin{equation}
    \small
    \label{Encoder_DisMatrix}
    \begin{split}
        \textbf{D}^t &= [d_{\mathit{v}_i^t, \mathit{v}_j^t}]_{1\leq i,j \leq |o|}, \\
        d_{\mathit{v}_i^t, \mathit{v}_j^t} &= sim(\textbf{v}_i^t,\textbf{v}_j^t), 
    \end{split}
\end{equation}
where $\textbf{D}^t \in \mathbb{R}^{|o|\times |o|}$ denotes the relation matrix, and $sim(\cdot)$ stands for similarity function (e.g., cosine similarity). Subsequently, for any student node $\textit{v}_i^t$, we select top-k other student nodes with the highest potential associations to construct its first-order neighbors according to $\textbf{D}^t_{i,j}, j\in \{1,\ldots,|o|\} \setminus \{i\}$, and then we can get the adjacency matrix $\textbf{A}^t \in \{0,1\}^{(|o|+1)\times(|o|+1)}$, which includes the connectivity between group node and student node, as shown in upper part of Figure~2(b).

After the construction of $\mathcal{G}^o$, we design a dynamic GCN module composed of static GCN units to effectively model the deeper associations between group and student as well as student and student by introducing the GCN layers~\cite{GCN}. Given the node feature matrix $\textbf{V}^t$ and the adjacency matrix $\textbf{A}^t$, the $l$-layer GCN of the unit under the $t$-th time frame is defined as:
\begin{equation}
    \small
    \label{Encoder_GCN}
    \begin{split}
        \textbf{V}^t_{(l+1)} = \alpha(\hat{\textbf{A}}^t \textbf{V}^t_{(l)} \textbf{W}^t_{(l)} + \textbf{b}^t_{(l)}),
    \end{split}
\end{equation}
where ${\textbf{V}^t_0} = \textbf{V}^t$,  ${\textbf{W}^t_{(l)}} \in \mathbb{R}^{d_{l}\times d_{l+1}}$, ${\textbf{b}^t_{(l)}} \in \mathbb{R}^{d_l}$ are trainable parameters, $d_l$ denotes the output node feature dimension for $l$-th layer and $d_0 = 2d$, $\alpha(\cdot)$ stands for the non-linear activation function (e.g., ReLU), and $\hat{\textbf{A}}^t = (\tilde{\textbf{U}}^t)^{-\frac{1}{2}} \tilde{\textbf{A}}^t (\tilde{\textbf{U}}^t)^{-\frac{1}{2}})$ is the normalized symmetric adjacency matrix. Here, $\tilde{\textbf{A}}^t = \textbf{A}^t + \textbf{I}_{|o|+1}$, and $\tilde{\textbf{U}}^t$ is the degree matrix. $\textbf{I}_{|o|+1}$ denotes an identity matrix with dimensions of $|o|+1$ and $\tilde{\textbf{U}}^t_{ii} = \sum_j\textbf{A}^t_{ij}$.

\subsubsection{Temporal Self-Attentive Network}

To more effectively capture the intricate temporal effects in the overall abilities and knowledge states of groups and individual students during the learning process, we introduce a temporal self-attentive network in this section. Following previous work~\cite{AKT,simpleKT}, we define the temporal self-attention module and obtain the retrieved knowledge states of the group and student (i.e., the selection of $\mathcal{Q}, \mathcal{K}, \mathcal{V}$ parameters is essentially a knowledge retriever) as follows:
\begin{equation}
    \small
    \centering
    \label{Encoder_SelfAtt}
    \left\{
    \begin{aligned}
         \textbf{h}_{t+1}^o &= SelfAtt(\mathcal{Q}^o,\mathcal{K}^o,\mathcal{V}^o), \\
         \mathcal{Q}^o&={\textbf{m}}^{t+1}_0, \mathcal{K}^o=\{{\textbf{m}}^{1}_0,\ldots,{\textbf{m}}^{t}_0\}, \mathcal{V}^o=\{{\textbf{n}}^{1}_0,\ldots,{\textbf{n}}^{t}_0\}; \\
         \textbf{h}_{t+1}^{s_i} &= SelfAtt(\mathcal{Q}^{s_i},\mathcal{K}^{s_i},\mathcal{V}^{s_i}), \\
         \mathcal{Q}^{s_i}&={\textbf{m}}^{t+1}_i, \mathcal{K}^{s_i}=\{{\textbf{m}}^{1}_i,\ldots,{\textbf{m}}^{t}_i\}, \mathcal{V}^{s_i}=\{{\textbf{n}}^{1}_i,\ldots,{\textbf{n}}^{t}_i\}, \\
    \end{aligned}
    \right. 
\end{equation}
where ${\textbf{m}}^{t}_0 = {\textbf{V}^t_{(L)}}[0,:\!d], {\textbf{n}}^{t}_0 = {\textbf{V}^t_{(L)}}[0,d\!:], {\textbf{m}}^{t}_i = {\textbf{V}^t_{(L)}}[i,:\!d], {\textbf{n}}^{t}_i = {\textbf{V}^t_{(L)}}[i,d\!:] \in \mathbb{R}^{d}, 1\leq i \leq |o|$ are the interaction features of group $o$ and student $s_i$ extracted from the learned $L$-th layer node representations, $d = \frac{d_L}{2}$ is the dimension size of final representation and $SelfAtt(\cdot)$ denotes the Self-Attention module. Notably, similar to AKT~\cite{AKT}, we can only model current interaction with visible historical learning behaviors to prevent information leakage (as shown in the lower part of Figure~2(b)).


Finally, we construct a readout module consisting of a two-layer fully connected network for the next time frame performance prediction of the group and student (as shown in part (1) of Figure~2(c)). Specifically:
\begin{equation}
    \small
    \centering
    \label{Encoder_Readout}
    \left\{
    \begin{aligned}
        \hat{y}_{i,t+1} &= MLP_{O}(\textbf{[}\textbf{h}_{t+1}^o ;\textbf{e}_i^{t+1}\textbf{]}), \\
        \hat{r}_{i,t+1} &= MLP_{S}(\textbf{[}\textbf{h}_{t+1}^s ;\textbf{e}_i^{t+1}\textbf{]}),
    \end{aligned}
    \right. 
\end{equation}
where $\hat{y}_{i,t+1}$ and $\hat{r}_{i,t+1}$ denote the predicted responses of group $o$ and student $s$ on $i$-th exercise $\textbf{e}_i^{t+1}$ under $t\!+\!1$-th time frame, and $MLP_O$ and $MLP_S$ are two~MLP networks.

\vspace{-3mm}
\subsection{Model Optimization}

In this section, we describe the model training and parameter optimization. In particular, we design a contrastive learning module in order to ensure the stability of the training process and the effectiveness of the learning of representations.

\subsubsection{Contrastive Learning Module}
Inspired by~\cite{DTransformer,CL4KT}, we argue that a learner may have response biases, such as careless responses or guessing responses, when performing questions over a period of time. Along this line, we design a bias-aware contrastive learning module for learning robust representations. Specifically, as shown in part (2) of Figure~2(c), we randomly flip students' responses in the time frames to generate the augmented interactions, and intuitively, the augmented student state should remain similar to the original. Following previous work~\cite{DTransformer}, we calculate contrastive loss from a single batch:
\begin{equation}
    \small
    \centering
    \label{CL_loss}
    \begin{split}
        \mathcal{L}^{C\!L}_{s_i,t} = -log\frac{exp(sim(\textbf{h}^{s_i}_t, \textbf{h}^{s_i\!+}_t)/\tau)}{\sum_{j\ne i}exp(sim(\textbf{h}^{s_i}_t, \textbf{h}^{s_j\!+}_t)/\tau)},
    \end{split} 
\end{equation}
where $\textbf{h}^{s_i\!+}_t$ denotes the augmented student state of student $s_i$ under the $t$-th time frame, $\tau$ is a temperature parameter (here set as 0.05), and $sim(\cdot)$ stands for the similarity function.

\subsubsection{Loss Function}
In the training phase, we jointly evaluate the model performance by predicting the interaction of both the group- and student-exercise responses. Specifically, for each student $s$ within group $o$, we utilize the cross-entropy loss function for the student performance prediction:
\begin{equation}
    \small
    \label{Loss_student}
    \begin{split}
        \mathcal{L}_o^{stu} = - \sum\limits_{s}\sum\limits_{t} \sum\limits_{i} r^s_{i,t}\log\hat{r}^s_{i,t}-(1-r^s_{i,t})\log(1-\hat{r}^s_{i,t}).
    \end{split}
\end{equation}

\noindent Then, we choose the mean square error loss (MSE) function for the group-exercise interaction prediction: 
\begin{equation}
    \small
    \label{Loss_group}
    \begin{split}
        \mathcal{L}_o^{grp} = \sum\limits_{t}\sum\limits_{i}(\hat{y}^o_{i,t}-y^o_{i,t})^2.
    \end{split}
\end{equation}

\noindent Meanwhile, we calculate the contrastive loss of all students in the group $o$ according to Eq.(13):
\begin{equation}
    \small
    \label{Loss_CL}
    \begin{split}
        \mathcal{L}_o^{C\!L} = \sum\limits_{s}\sum\limits_{t} \mathcal{L}^{C\!L}_{s,t}.
    \end{split}
\end{equation}

\noindent Finally, we obtain the complete optimization objective function based on the above three loss objectives:
\begin{equation}
    \small
    \label{Loss}
    \begin{split}
        \mathcal{L}_o = \mathcal{L}_o^{grp} + \frac{1}{|o|} \; \mathcal{L}_o^{stu} + \gamma \cdot \mathcal{L}^{C\!L}_o.
    \end{split}
\end{equation}
where $\gamma$ is the weight coefficient to control the influence of contrastive signals. We can then train the whole model and optimize the model parameters utilizing gradient descent.




\section{Experiments}
In this section, we conduct extensive experiments on four real-world education datasets aiming at verifying the effectiveness and superiority of our proposed RIGL model. Specifically, we will answer the following research questions (\textbf{RQs}) to unfold the experiments:

\begin{itemize}[leftmargin=*]
    \item[$\bullet$] \textbf{RQ1:} What about the effectiveness and superiority of the proposed RIGL model on the holistic knowledge tracing task?
    %
    \item[$\bullet$] \textbf{RQ2:} Do the designed key components benefit our proposed RIGL model in achieving performance improvement? 
    \item[$\bullet$] \textbf{RQ3:} How do the hyper-parameter settings influence the holistic knowledge tracing performance of the RIGL model?
    \item[$\bullet$] \textbf{RQ4:} How does RIGL facilitate tracing the evolution of knowledge states in both individuals and groups and how does it help to understand the progression of their relationships over time?
    
\end{itemize}




\begin{table}[!t]
    \centering
    \caption{The statistics of all datasets.}
    \vspace{-2mm}
    \label{table.dataset_statistic}
    \resizebox{\linewidth}{!}{ 
        \Huge
        \begin{tabular}{lrrrr}
        \toprule
        Statistics & ASSIST12 & NIPS-Edu & SLP-Math & SLP-Bio\\
        \midrule
        \#Students & 2,281 & 1,138 & 1,488 & 1,727 \\
        \#Groups   & 101 & 91 & 126 & 150 \\
        \#Exercises & 8,838 & 747 & 142 & 121 \\
        \#Knowledge concepts & 162 & 225 & 41 & 22\\
        Avg. group size & 22.79 & 12.51 & 11.80 & 11.51\\
        Avg. responses per student & 67.16 & 87.53 & 78.54 & 94.58 \\
        Avg. responses per group & 55.33 & 67.46 & 79.93 & 97.84 \\
        Avg. responses per time frame & 33.53 & 32.47 & 32.93 & 35.04 \\
        \bottomrule
        \end{tabular}
    }
    \vspace{-5mm}
\end{table}

\vspace{-3mm}
\subsection{Experimental Setting}

\subsubsection{\textbf{Datasets.}}  We conduct experiments on four real-world education datasets with diversity to evaluate the effectiveness of the proposed RIGL on the holistic tracing task, which are ASSIST12~\cite{ASSISTment12}, NIPS-Edu~\cite{NIPS-Edu}, SLP-Math~\cite{SLP} and SLP-Bio~\cite{SLP}. All datasets contain the group labels (i.e., the class to which the students belong), and students from the same group share the same label category. Table~1 shows the statistics of the datasets, and more details of the dataset description and data preprocessing are available in Appendix~A.2. 

\vspace{-1mm}
\subsubsection{\textbf{Baseline Approaches.}} The performance of RIGL is compared with eight strong and commonly used baselines including DKT~\cite{DKT}, SAKT~\cite{SAKT}, AKT~\cite{AKT}, LPKT~\cite{LPKT}, GIKT~\cite{GIKT}, simpleKT~\cite{simpleKT}, AT-DKT~\cite{AT_DKT} and DTransformer~\cite{DTransformer}. Notably, these baselines are individual-based knowledge tracing models, so we adapt them on the HKT task. The introduction and implementation details can be found in Appendix~A.3.

\vspace{-1mm}
\subsubsection{\textbf{Evaluation Metrics.}}
To comprehensively evaluate the performance of all methods on holistic knowledge tracing, we adopt four evaluation metrics, including the area under the ROC curve (AUC), accuracy~(ACC), root mean square error~(RMSE), and mean absolute error~(MAE). Specifically, the AUC and ACC are used to evaluate the individual-level performance prediction, and RMSE and MAE for the group-level performance on future exercises.

\begin{table*}[!t]
    \setlength{\tabcolsep}{1.5pt}
    \renewcommand{\arraystretch}{1.1}
    \centering
    \vspace{-3mm}
    \caption{Performance of RIGL and all baselines on all datasets on predicting future performance of individual-level and group-level. $\uparrow$ ($\downarrow$) means the higher (lower) score the better performance. “$*$” denotes the statistically significant improvement of RIGL model compared to the best baseline method (i.e., two-sided t-test with \textit{p}<0.05). $\bold{Bold}$: the best, $\underline{\rm Underline}$: the runner-up.}
    \vspace{-3mm}
    \label{table.Result_Table}
    \resizebox{1.0\linewidth}{!}{ 
        \small
        \begin{tabular}{l|cc|cc|cc|cc}
            \toprule
            \multirow{2}{*}{Datasets} & \multicolumn{4}{|c}{ASSIST12} & \multicolumn{4}{|c}{NIPS-Edu}  \\
            \cmidrule{2-9} 
            \multicolumn{1}{c|}{} & \multicolumn{2}{|c}{Individual-level} & \multicolumn{2}{|c}{Group-level} & \multicolumn{2}{|c}{Individual-level} & \multicolumn{2}{|c}{Group-level}  \\
            \midrule
            Metrics & AUC ($\uparrow$) & ACC ($\uparrow$) & RMSE ($\downarrow$) & MAE ($\downarrow$) & AUC ($\uparrow$) & ACC ($\uparrow$) & RMSE ($\downarrow$) & MAE ($\downarrow$)  \\
            \midrule
            DKT & $0.6276_{\pm 0.0282}$ & $0.7042_{\pm 0.0174}$ & $0.2331_{\pm 0.0164}$ & $0.1776_{\pm 0.0120}$ & $0.6267_{\pm 0.0144}$ & $0.6159_{\pm 0.0219}$ & $0.2957_{\pm 0.0034}$ & $0.2373_{\pm 0.0026}$ \\
            SAKT & $0.6075_{\pm 0.0235}$ & $0.7183_{\pm 0.0232}$ & $0.2302_{\pm 0.0162}$ & $0.1693_{\pm 0.0075}$ & $0.6295_{\pm 0.0183}$ & $0.6207_{\pm 0.0164}$ & $0.3060_{\pm 0.0143}$ & $0.2470_{\pm 0.0126}$ \\
            AKT & $0.6188_{\pm 0.0217}$ & $0.7259_{\pm 0.0245}$ & $0.2288_{\pm 0.0205}$ & $0.1750_{\pm 0.0113}$ & $0.6448_{\pm 0.0159}$ & $0.6284_{\pm 0.0259}$ & $0.2768_{\pm 0.0022}$ & $0.2275_{\pm 0.0048}$ \\
            LPKT & $0.6379_{\pm 0.0287}$ & $0.7362_{\pm 0.0300}$ & $0.2194_{\pm 0.0205}$ & $0.1705_{\pm 0.0111}$ & $0.6630_{\pm 0.0135}$ & $0.6453_{\pm 0.0129}$ & $\underline{0.2665_{\pm 0.0073}}$ & $0.2185_{\pm 0.0074}$ \\
            GIKT & $0.6268_{\pm 0.0201}$ & $0.7271_{\pm 0.0212}$ & $0.2259_{\pm 0.0133}$ & $0.1759_{\pm 0.0115}$ & $0.6547_{\pm 0.0136}$ & $0.6410_{\pm 0.0227}$ & $0.2744_{\pm 0.0094}$ & $0.2431_{\pm 0.0105}$ \\
            simpleKT & $0.6161_{\pm 0.0214}$ & $0.7295_{\pm 0.0255}$ & $0.2226_{\pm 0.0182}$ & $\underline{0.1654_{\pm 0.0124}}$ & $0.6402_{\pm 0.0206}$ & $0.6268_{\pm 0.0230}$ & $0.2736_{\pm 0.0136}$ & $0.2179_{\pm 0.0112}$ \\
            AT-DKT & $0.6414_{\pm 0.0223}$ & $0.7395_{\pm 0.0197}$ & $0.2212_{\pm 0.0174}$ & $0.1728_{\pm 0.0106}$ & $0.6678_{\pm 0.0203}$ & $\underline{0.6525_{\pm 0.0233}}$ & $0.2683_{\pm 0.0087}$ & $0.2232_{\pm 0.0096}$ \\
            DTransformer & $0.6392_{\pm 0.0237}$ & $0.7347_{\pm 0.0216}$ & $\underline{0.2183_{\pm 0.0157}}$ & $0.1686_{\pm 0.0097}$ & $\underline{0.6726_{\pm 0.0113}}$ & $0.6488_{\pm 0.0137}$ & $0.2706_{\pm 0.0074}$ & $\underline{0.2163_{\pm 0.0122}}$ \\
            \midrule
            LPKT-Ind & $\underline{0.7344_{\pm 0.0039}}$ & $\underline{0.7579_{\pm 0.0067}}$ & - & - & $0.6541_{\pm 0.0061}$ & $0.6296_{\pm 0.0138}$ & - & - \\
            simpleKT-Ind & $0.6696_{\pm 0.0073}$ & $0.7481_{\pm 0.0074}$ & - & - & $0.6425_{\pm 0.0032}$ & $0.6287_{\pm 0.0076}$ & - & - \\
            \midrule
            \midrule
            RIGL & $\bm{0.7394^*_{\pm 0.0141}}$ & $\bm{0.7673^*_{\pm 0.0126}}$ & $\bm{0.2074^*_{\pm 0.0103}}$ & $\bm{0.1515^*_{\pm 0.0109}}$ & $\bm{0.7326^*_{\pm 0.0115}}$ & $\bm{0.6779^*_{\pm 0.0144}}$ & $\bm{0.2344^*_{\pm 0.0093}}$ & $\bm{0.1775^*_{\pm 0.0072}}$ \\
            \bottomrule
        \end{tabular}
    }
    \resizebox{1.0\linewidth}{!}{   
        \small
        \begin{tabular}{l|cc|cc|cc|cc}
            \toprule
            \multirow{2}{*}{Datasets} & \multicolumn{4}{|c}{SLP-Math} & \multicolumn{4}{|c}{SLP-Bio}  \\
            \cmidrule{2-9} 
            \multicolumn{1}{c|}{} & \multicolumn{2}{|c}{Individual-level} & \multicolumn{2}{|c}{Group-level} & \multicolumn{2}{|c}{Individual-level} & \multicolumn{2}{|c}{Group-level}  \\
            \midrule
            Metrics & AUC ($\uparrow$) & ACC ($\uparrow$) & RMSE ($\downarrow$) & MAE ($\downarrow$) & AUC ($\uparrow$) & ACC ($\uparrow$) & RMSE ($\downarrow$) & MAE ($\downarrow$)  \\
            \midrule
            DKT & $0.7765_{\pm 0.0036}$ & $0.7636_{\pm 0.0159}$ & $0.3136_{\pm 0.0165}$ & $0.2439_{\pm 0.0166}$ & $0.7152_{\pm 0.0050}$ & $0.7147_{\pm 0.0159}$ & $0.1975_{\pm 0.0035}$ & $0.1580_{\pm 0.0030}$ \\
            SAKT & $0.7899_{\pm 0.0033}$ & $0.7635_{\pm 0.0102}$ & $0.2045_{\pm 0.0098}$ & $0.1691_{\pm 0.0089}$ & $0.7366_{\pm 0.0043}$ & $0.7151_{\pm 0.0094}$ & $0.2167_{\pm 0.0050}$ & $0.1759_{\pm 0.0048}$ \\
            AKT & $\underline{0.7957_{\pm 0.0060}}$ & $0.7742_{\pm 0.0110}$ & $0.1895_{\pm 0.0072}$ & $0.1559_{\pm 0.0074}$ & $0.6850_{\pm 0.0097}$ & $0.6923_{\pm 0.0131}$ & $0.2122_{\pm 0.0037}$ & $0.1749_{\pm 0.0055}$ \\
            LPKT & $0.7844_{\pm 0.0093}$ & $0.7249_{\pm 0.0235}$ & $0.1947_{\pm 0.0038}$ & $0.1569_{\pm 0.0026}$ & $0.7210_{\pm 0.0042}$ & $0.6754_{\pm 0.0146}$ & $\underline{0.1966_{\pm 0.0034}}$ & $\underline{0.1561_{\pm 0.0028}}$ \\
            GIKT & $0.7818_{\pm 0.0087}$ & $0.7643_{\pm 0.0154}$ & $0.2175_{\pm 0.0069}$ & $0.1624_{\pm 0.0065}$ & $0.7223_{\pm 0.0073}$ & $0.7017_{\pm 0.0048}$ & $0.2082_{\pm 0.0051}$ & $0.1710_{\pm 0.0066}$ \\
            simpleKT & $0.7904_{\pm 0.0084}$ & $0.7725_{\pm 0.0095}$ & $0.2038_{\pm 0.0129}$ & $\underline{0.1536_{\pm 0.0133}}$ & $0.6919_{\pm 0.0051}$ & $0.6958_{\pm 0.0117}$ & $0.2040_{\pm 0.0029}$ & $0.1650_{\pm 0.0032}$ \\
            AT-DKT & $0.7865_{\pm 0.0047}$ & $0.7691_{\pm 0.0081}$ & $0.1906_{\pm 0.0113}$ & $0.1577_{\pm 0.0085}$ & $0.7064_{\pm 0.0118}$ & $0.6976_{\pm 0.0085}$ & $0.2038_{\pm 0.0063}$ & $0.1685_{\pm 0.0049}$ \\
            DTransformer & $0.7919_{\pm 0.0043}$ & $\underline{0.7757_{\pm 0.0095}}$ & $\underline{0.1874_{\pm 0.0064}}$ & $0.1546_{\pm 0.0028}$ & $0.7367_{\pm 0.0069}$ & $\underline{0.7173_{\pm 0.0104}}$ & $0.1981_{\pm 0.0075}$ & $0.1597_{\pm 0.0050}$ \\
            \midrule
            LPKT-Ind & $0.7762_{\pm 0.0086}$ & $0.7387_{\pm 0.0179}$ & - & - & $\underline{0.7375_{\pm 0.0018}}$ & $0.7149_{\pm 0.0083}$ & - & - \\
            simpleKT-Ind & $0.7816_{\pm 0.0057}$ & $0.7566_{\pm 0.0169}$ & - & - & $0.6961_{\pm 0.0162}$ & $0.7099_{\pm 0.0127}$ & - & - \\
            \midrule
            \midrule
            RIGL & $\bm{0.8304^*_{\pm 0.0049}}$ & $\bm{0.7853^*_{\pm 0.0102}}$ & $\bm{0.1383^*_{\pm 0.0048}}$ & $\bm{0.1078^*_{\pm 0.0113}}$ & $\bm{0.7959^*_{\pm 0.0080}}$ & $\bm{0.7442^*_{\pm 0.0085}}$ & $\bm{0.1357^*_{\pm 0.0042}}$ & $\bm{0.1058^*_{\pm 0.0039}}$ \\
            \bottomrule
        \end{tabular}
    }
    \vspace{-3mm}
    
\end{table*}

\begin{figure*}[!t] 
    \vspace{-3mm}
	\centering 
	\includegraphics[scale=0.49]{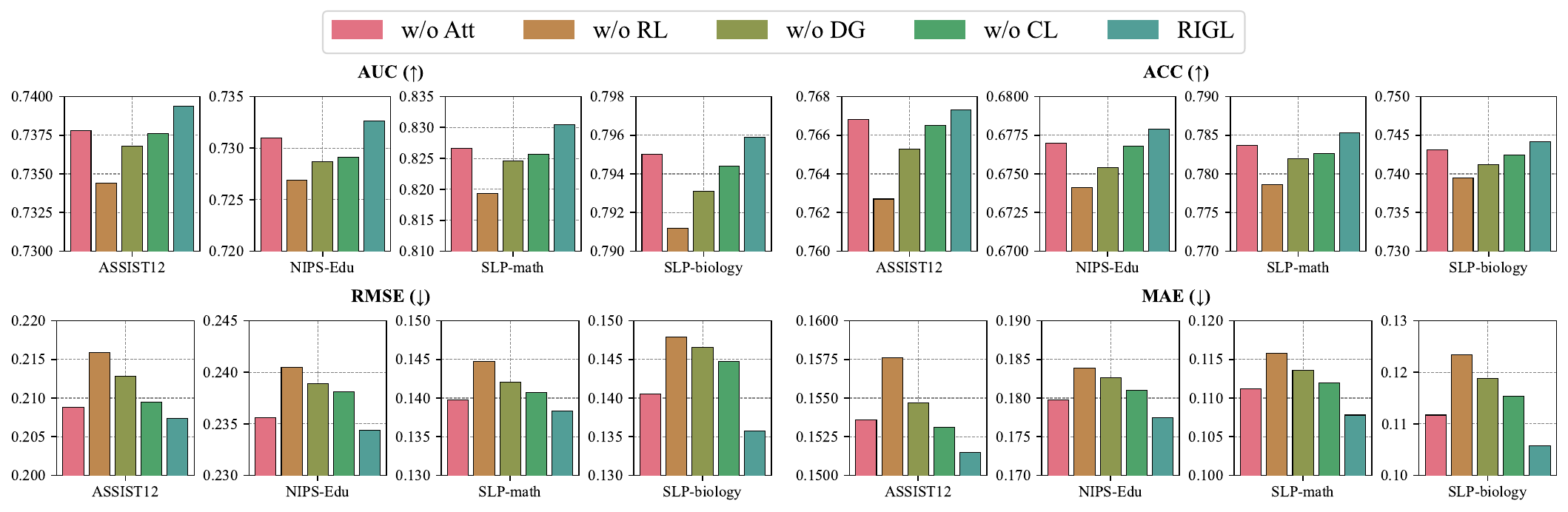} 
    \vspace{-4mm}
	\caption{Performance of ablation studies conducted on four datasets, where “w/o” means removing the target module.}
    \label{fig.ablition1} 
    \vspace{-3mm}
\end{figure*}

\vspace{-1mm}
\subsubsection{\textbf{Parameter Settings.}}
We performed $5$-fold cross-validation in the experiments. For each fold, 80\% of samples are set as for the training data, and others are for the test set. We implemented all methods with PyTorch by Python. The dimension size of embeddings (i.e., $d_e$, $d_c$, $d_r$) was set as 256. The number of the GCN layers $L_1$ was set to 2, where each layer's hidden size is 256, and the number of self-attentive network layers $L_2$ is set to 4. We used the Adam optimizer, where the learning rate was searched in [1e-4, 5e-4, 1e-3, 2e-3, 1e-2]. The coefficient $\gamma$ of contrastive loss was set to 0.01. We adopted the cosine similarity as the similarity calculation function $sim(\cdot)$~(Eq.9~and~Eq.13).

\begin{figure}[!t] 
	\centering 
	\includegraphics[scale=0.46]{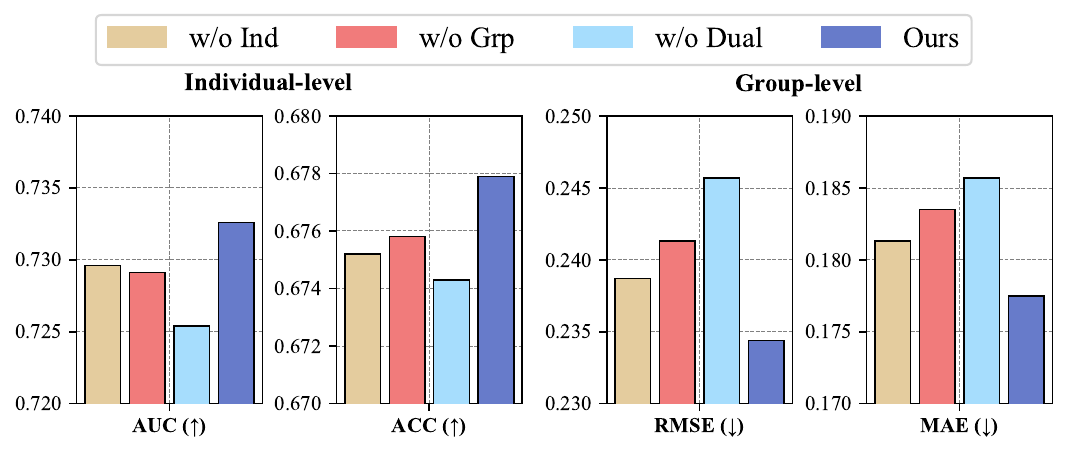} 
    \vspace{-3mm}
	\caption{Results of reciprocal effect study conducted on NIPS-Edu dataset, where “w/o” means removing the target feature.}
    \label{fig.case_study_1} 
    \vspace{-6mm}
\end{figure}


\vspace{-3mm}
\subsection{Performance Comparison (RQ1)}

Table~2 shows the performance of the proposed RIGL including individual-level and group-level compared with the baseline models on the four datasets. We highlighted the best results of all models in boldface and underlined the suboptimal results. According to the results, there are several observations: (1) Our RIGL model demonstrates significant improvements over the baselines across all datasets. Specifically, it shows an average increase of 4.01\% and 20.32\% over the best baseline model at both the individual and group levels, respectively, which underscores the effectiveness of the RIGL. Especially, in comparison to the runner-up method on the SLP-Bio dataset, our model achieves an average of 5.83\% and 31.59\% performance improvements in terms of individual-level and group-level. (2) RIGL generally has a higher percentage of increased performance in terms of the group-level than the individual-level, demonstrating that personalized learning information is highly effective and has a more pronounced impact on the group-level diagnosis during the reciprocal learning modeling. (3) In the baseline methods, the DTransformer and LPKT models exhibit relatively better performance compared to other baselines. This phenomenon may be attributed to the gains from modeling forgetting behaviors in both group and individual learning processes in those two models. 

In particular, we also conducted experiments for baselines' individual version (i.e., traditional knowledge tracing model) utilizing only individual interaction data and present the results of LPKT-Ind and simpleKT-Ind in Table~2 (the complete results can be found in Appendix~A.4). As can be observed from the results, our RIGL remains significantly superior in terms of individual-level proficiency assessment. Specifically, it shows an average increase of 5.91\% and 3.48\% over the best individual knowledge tracing baseline in terms of AUC and ACC metrics, respectively, which demonstrates the contribution of group learning behaviors to the modeling of independent learning, and the effectiveness of the proposed reciprocal learning modeling in our RIGL model. In addition, some of the baseline models are less effective at the individual-level than modeling the individual alone, suggesting that simple joint training may not always work, and this is further evidence of the validity of RIGL. An interesting phenomenon is that the advantage of RIGL over the KTM-ind model behaves differently on different datasets, which may be caused by the characteristics of the dataset. Since the RIGL model requires tracing both individual and group ability changes, its performance suffers when the group-exercise interactions are relatively sparse in the dataset. Whereas KTM-ind (e.g., LPKT-ind) as the primitive individual KT model is unable to model group interactions (focusing on individual level modeling) and thus is less affected by this sparsity.

\subsection{Ablation Study (RQ2)}
To answer RQ2, we first conducted a comprehensive ablation study to investigate the impact of each module in the RIGL model by defining the following variations: 1) \textbf{w/o Att}: removing the absence-perceived attention aggregation module in reciprocal enhanced learning; 2) \textbf{w/o RL}: removing the reciprocal enhanced learning module; 3) \textbf{w/o DG}: removing the dynamic graph modeling; 4) \textbf{w/o CL}: removing the contrastive loss. As illustrated in Figure~3, the results reveal insightful observations: (1) In comparison to RIGL, all variants suffer relative performance degradation on four datasets across various evaluation metrics, demonstrating the contribution of the designed submodules to our proposed model. (2) The most significant decrease in the model performance occurs after removing the reciprocal enhanced learning module, which exhibits that jointly and interactively modeling the reciprocal learning process of students and groups plays a crucial role in the HKT task, and also confirms the soundness of our designed model. (3) The performance degradation of removing the dynamic graph modeling module is also quite noticeable, reflecting the fact that dynamic graph modeling is important for capturing the evolving knowledge states of groups and individuals.

\begin{figure}[!t] 
    \vspace{-1mm}
	\centering 
	\includegraphics[scale=0.37]{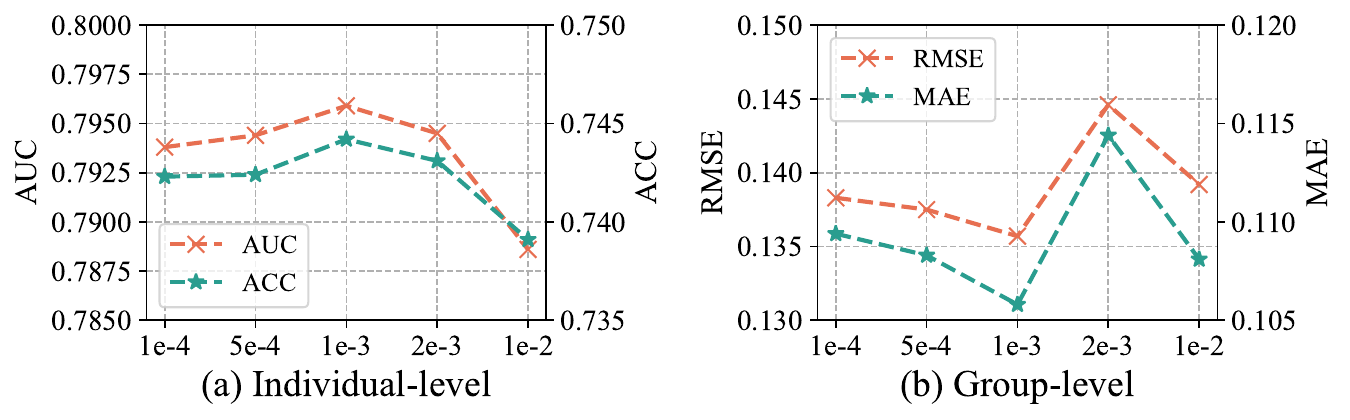} 
    \vspace{-8mm}
	\caption{Sensitivity analysis of learning rate on SLP-Bio.}
    \label{fig.para_study_1} 
    \vspace{-3mm}
\end{figure}

\begin{figure}[!t] 
    \vspace{-1mm}
	\centering 
	\includegraphics[scale=0.37]{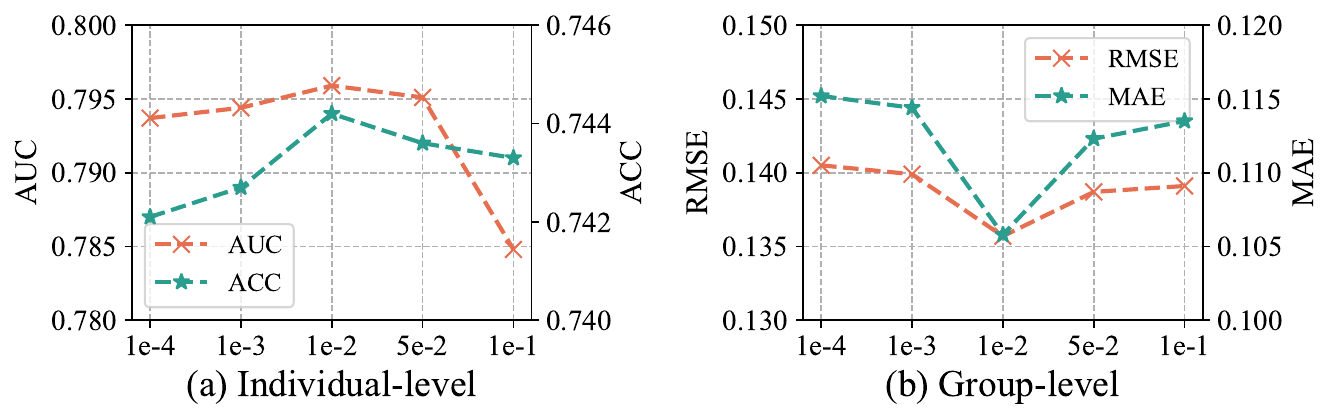} 
    \vspace{-4mm}
	\caption{Sensitivity analysis of coefficient $\gamma$ on SLP-Bio.}
    \label{fig.para_study_2} 
    \vspace{-5mm}
\end{figure}

\begin{figure*}[!t] 
    \vspace{-1mm}
	\centering 
	\includegraphics[scale=1.13]{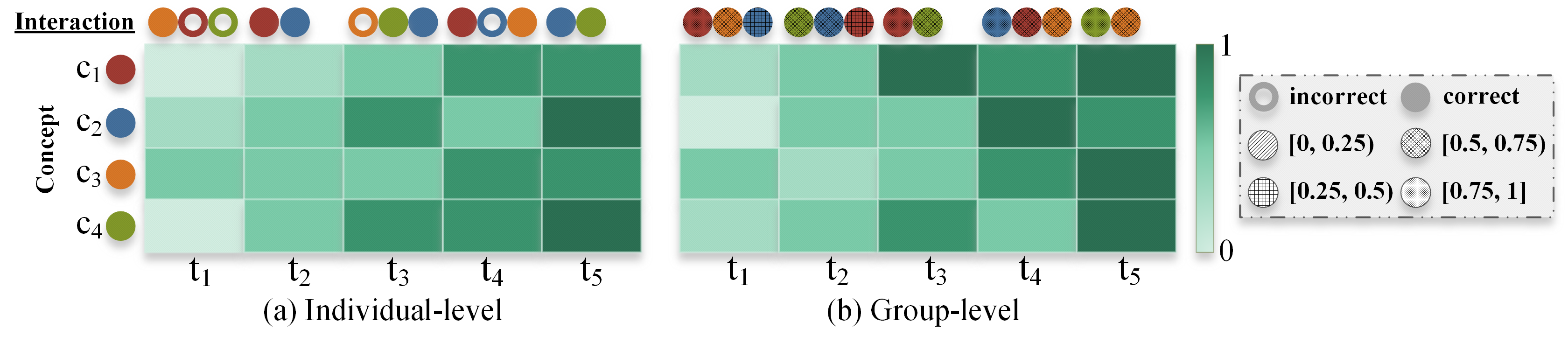} 
    \vspace{-1mm}
	\caption{The evolution process of the individual-level and group-level knowledge proficiency on four concepts during the holistic knowledge tracing traced by RIGL. The four knowledge concepts are depicted in different colors and the top line represents the exercises answered by the learner and the group in each time frame (different colors denote corresponding to different concepts). In addition, solid and hollow circles indicate correct and incorrect learner responses, respectively, as well as circles with different spots denote different ranges of response values for the group.}
    \vspace{-1mm}
    \label{fig.case_study_2} 
\end{figure*}

In addition, we further conducted an ablation study to investigate the effect brought by reciprocal learning on the performance of holistic knowledge tracing by disassembling the reciprocal enhanced learning module in RIGL. Specifically, we removed the individual-level features, the group-level features, and both features in the reciprocal enhanced learning module (i.e., \textbf{w/o Ind}, \textbf{w/o Grp}, and \textbf{w/o Dual}), respectively, and then inspected the performance variations. Notably, the experiments in this section are completed by removing the features of individual and group interactions in the reciprocal module, respectively, within the framework of HKT, which requires that the ablated RIGL is still capable of tracking both the evolution of individual and group abilities. The experimental results are shown in Figure~4. We have the following observations: (1) The absence of individual-level and group-level features brings about performance degradation, demonstrating the complementary facilitation of individuals' and groups' learning features in the HKT task. (2) The performance degradation introduced when neither individual modeling nor group modeling utilizes each other's features during the information fusion process is very significant, which is further evidence of the effectiveness of reciprocal learning. 

\vspace{-2mm}
\subsection{Parameter Sensitivity Analysis (RQ3)}
To answer RQ3, we conducted a parameter sensitivity analysis in this section to investigate the effects of hyper-parameters, which mainly include the learning rate and the weight coefficient $\gamma$ of the contrastive loss. Specifically, we set the list of learning rates to be \{1e-4, 5e-4, 1e-3, 2e-3, 1e-2\}, as well as the $\gamma$ values \{1e-4, 1e-3, 1e-2, 5e-2, 1e-1\}, and mainly show the experimental results on the SLP-Bio dataset. As shown in Figure~5, we observe that 1e-3 is sufficient for the learning rate, and the performance presents a trend of rising first and then falling with the increase of the rate. As shown in Figure~6, the model reaches its best performance when the value of $\gamma$ is 1e-2 and the effect on the AUC metric is not very strong. An interesting phenomenon here is that the trend of this coefficient's impact on the performance of the individual-level and the group-level is different as the $\gamma$ value increments, which is perhaps due to the different modeling effects of the fraction of contrastive loss on group learning and individual learning behaviors.

\subsection{Case Study (RQ4)}


\begin{figure}[!t] 
	\centering 
	\includegraphics[scale=0.89]{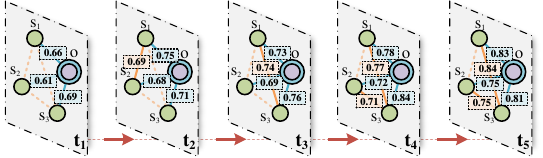} 
    \vspace{-1mm}
	\caption{Case study of the dynamic relationship mining. Different colored lines represent different relationships, and dashed lines indicate connections below the threshold. The learned edge weights denote the similarity, i.e., higher values denote greater similarity between the two.}
    \label{fig.ablition2} 
    \vspace{-3mm}
\end{figure}

\subsubsection{\textbf{Visualization of Proficiency Evolution.}} To further understand how RIGL traces the evolution of the individual-level and group-level knowledge states, in this case study, we demonstrated the tracing process. Figure~7 illustrates the evolution of both a learner's and a group's knowledge proficiency across five time frames on four knowledge concepts, as traced by RIGL. From the visualization, we can observe that the individual's mastery of the corresponding concept of the exercise increases even if it is answered incorrectly, which implies that wrongly responding to the exercise also brings knowledge gain. In addition, it can be seen that the knowledge gained from individual learning affects the group's knowledge level, yet the degree of impact varies under different time frames, meaning that the effect of the individual on the group varies over~time, which is in line with the real scenario.

\subsubsection{\textbf{Dynamic Relationship Mining.}} Specifically, a case study of relationship observations was conducted to explore the dynamic relationships between individuals and the group, as well as among individuals in the holistic knowledge tracing process. Figure~8 shows the dynamic relationship graph across the group under five time frames by presenting the correlation distances (as mentioned in Section~4.2; for convenience of the display, three students under one group were selected). It can be observed that the relationships between students and groups and among students are dynamically changing (e.g., the increasing similarity of both $s_1$ and $o$ as well as $s_1$ and $s_3$), justifying the exploitation and modeling of potential associations through our use of the dynamic graph. In addition, based on the mined latent relationships, potential subgroups within the group can be effectively identified, which will help us to perceive the evolution of the internal structure of the group.

\section{Conclusion}
In this paper, we introduced a novel framework termed as \textbf{RIGL} (a unified \underline{\textbf{R}}eciprocal approach for \underline{\textbf{I}}ndependent and \underline{\textbf{G}}roup \underline{\textbf{L}}earning processes), which aims to provide comprehensive and dynamic modeling for both independent learning and group learning. Our approach comprises several key components. Initially, we devised a time frame-aware reciprocal embedding module to concurrently capture the temporal interactions between students and groups during the learning processes, followed by a reciprocal-enhanced learning mechanism that maximizes the synergistic insights from these two learning behaviors. To further mine the intricate dynamics of student-group associations, we proposed a relation-guided temporal attentive network encompassing dynamic graph modeling and temporal self-attention mechanisms. Notably, the relation-guided dynamic graph is formed by uncovering potential links between students and groups. Finally, we incorporated a bias-aware contrastive learning module to ensure model stability during training. Extensive experiments were conducted on four real-world educational datasets to substantiate the efficacy of our RIGL model, particularly in addressing the HKT task. We hope this work could lead to further studies on holistic knowledge tracing.

\begin{acks}
This work was supported in part by the National Natural Science Foundation of China under Grant (No. U21A20512, No. 62107001, and No.62302010), in part by the Anhui Provincial Natural Science Foundation (No. 2108085QF272), and in part by by China Postdoctoral Science Foundation (No. 2023M740015).
\end{acks}

\newpage
\bibliographystyle{ACM-Reference-Format}
\bibliography{references}

\vspace{-3mm}
\appendix
\section{APPENDICES}

\renewcommand{\thetable}{S\arabic{table}}

\subsection{Notations}
To ensure clarity and comprehension for readers, the import notations used in this paper are meticulously presented in Table~\ref{table.math_notation}.

\setcounter{table}{0}
\begin{table}[!t]
    \caption{Summary of the primary notations.}
    \vspace{-2mm}
    \centering
    \begin{tabular}{c|p{5.3cm}}
        \hline
        \hline
         Symbols & Description \\
        \hline
        
        $\mathcal{O}, \mathcal{S}, \mathcal{E}, \mathcal{C}, \mathcal{Q}$ & The set of groups, students, exercises, knowledge concepts, and the Q-matrix, respectively. \\

        $o, s, e, c$ & The group, the student, the exercise, and the knowledge concept. \\

        $\mathcal{R}^S, \mathcal{R}^O$ & The whole interaction sequence of the student and the group, respectively.   \\

        $\mathcal{F}_t, \mathcal{H}_t$ & The student-exercise and group-exercise interaction sequence under $t$-th time frame. \\

        $r_i^t, y_i^t$ & The score that the student got on the $i$-th log under $t$-th time frame and the correct rate that group got on the $i$-th log under $t$-th time frame. \\ 

        \hline

        $\textbf{x}^t, \textbf{z}^t$ & The set of exercise encoding and response encoding, respectively. \\ 

        $\hat{\textbf{x}}^t, \hat{\textbf{z}}^t$ & The set of interaction encodings of groups. \\ 

        $\widetilde{\textbf{x}}_t^o, \widetilde{\textbf{z}}_t^o, \widetilde{\textbf{x}}_t^s, \widetilde{\textbf{z}}_t^s$ & The enhanced interaction encoding of group  and student. \\

        $\mathcal{G}_t^o = (V, E^t)$ & The group-individual graph with a node set $V$ and a edge set $E^t$. \\

        $\mathcal{V} = \{o, s_1, \ldots,s_{|o|}\} $ & The node set. \\

        $E^t = (E_{o \leftrightarrow s}, E_{s \leftrightarrow s}^t)$ & The edge set including the connecting edges between students and groups. \\

        $\textbf{V}^t, \textbf{D}^t, \textbf{A}^t$ & The node feature matrix, the relation matrix and the adjacency matrix in the $t$-th time frame, respectively. \\

        \makecell{$\mathcal{Q}^o,\mathcal{K}^o, \mathcal{V}^o$ \\ $\mathcal{Q}^s, \mathcal{K}^s, \mathcal{V}^s$} & The query matrix, the key matrix, and the value matrix of the self-attention module. \\

        $\textbf{m}, \textbf{n}$ & The interaction feature of the group and student extracted from the learned graph layer. \\
        
        $\textbf{h}_t^o, \textbf{h}_t^s$ & The knowledge states of group $o$ and student $s$ under $t$-the time frame, respectively. \\

        $\textbf{W}, \textbf{b}$ & The trainable matrix and parameters. \\

        \hline
        $|\cdot|$ & The cardinality of a set. \\

        $\tau, \gamma$& The temperature parameter and the coefficient weight parameter.\\ 
        
        $I, N, M, K$ & The size of the group set $\mathcal{O}$, the student set $\mathcal{S}$, the exercise set $\mathcal{E}$, and the concept set $\mathcal{C}$, respectively. \\

        $d_e, d_c, d_r, d_l, d$ & The dimension of exercise, the dimension of concept, the dimension of response, the dimension of $l$-th graph layer, and the hidden dimension, respectively. \\
        
        $\mathcal{L}_o, \mathcal{L}_o^{grp}, \mathcal{L}_o^{stu}, \mathcal{L}_o^{CL}$ & The total loss, the group loss, the student loss, and the contrastive loss, respectively. \\

        \hline
        \hline
    \end{tabular}
    \label{table.math_notation}
    \vspace{-5mm}
\end{table}

\subsection{Dataset Description and Preparation}
\subsubsection{\textbf{Data Description.}} In the experiments, we use four real-world education datasets to valid the effectiveness of our RIGL on the holistic tracing task, and the details are described as follows: 
\begin{itemize}[leftmargin=*]

    \item[$\bullet$] \textbf{ASSIST12}~\cite{ASSISTment12}: ASSIST12 is a dataset collected from the ASSISTments online educational tutoring system and contains extensive student exercise-solving data for the school year 2012-2013, which has been widely utilized in knowledge tracing tasks.

    \item[$\bullet$] \textbf{NIPS-Edu}~\cite{NIPS-Edu}: NIPS-Edu dataset was collected from the NeurIPS 2020 Education Challenge and contains records of students' responses to math exercises which include timestamp information.

    \item[$\bullet$] \textbf{SLP-Math}~\cite{SLP}: SLP is a public benchmark collected from an online learning platform called Smart Learning Partner (SLP), which intentionally records the learning data of secondary school students about multiple subjects to provide wealthy contents. SLP-Math is a sub-dataset of SLP on the subject of math.
    

    \item[$\bullet$] \textbf{SLP-Bio}~\cite{SLP}: Similar to SLP-Math, SLP-Bio is also a sub-dataset of SLP, which corresponds to the exercise-solving records of the secondary school students in the subject of biology.
\end{itemize}

\setcounter{table}{1}
\begin{table*}[!t]
    \vspace{-1mm}
    \setlength{\tabcolsep}{1.5pt}
    \renewcommand{\arraystretch}{1.1}
    \centering
    \caption{Performance of RIGL and baselines (only using individual data) on all datasets on predicting individual-level performance. $\uparrow$ ($\downarrow$) means the higher (lower) score the better performance. “$*$” denotes the statistically significant improvement of RIGL model compared to the best baseline method (i.e., two-sided t-test with \textit{p}<0.05). $\bold{Bold}$: the best, $\underline{\rm Underline}$: the runner-up.}
    \vspace{-3mm}
    \label{table.Result_Table}
    \resizebox{1.0\linewidth}{!}{ 
        \small
        \begin{tabular}{l|cc|cc|cc|cc}
            \toprule
            Datasets & \multicolumn{2}{|c}{ASSIST12} & \multicolumn{2}{|c|}{NIPS-Edu} &
            \multicolumn{2}{|c}{SLP-Math} & \multicolumn{2}{|c}{SLP-Bio} \\
            \midrule
            Metrics & AUC ($\uparrow$) & ACC ($\uparrow$) & AUC ($\uparrow$) & ACC ($\uparrow$) & AUC ($\uparrow$) & ACC ($\uparrow$) & AUC ($\uparrow$) & ACC ($\uparrow$) \\
            \midrule
            DKT-Ind & $0.7191_{\pm 0.0030}$ & $0.7440_{\pm 0.0051}$ & $0.6448_{\pm 0.0060}$ & $0.6228_{\pm 0.0096}$ & $0.7975_{\pm 0.0052}$ & $0.7689_{\pm 0.0092}$ & $0.7141_{\pm 0.0032}$ & $0.6678_{\pm 0.0159}$ \\
            SAKT-Ind & $0.6996_{\pm 0.0046}$ & $0.7489_{\pm 0.0048}$ & $0.6526_{\pm 0.0060}$ & $\underline{0.6304_{\pm 0.0109}}$ & $\underline{0.8060_{\pm 0.0037}}$ & $\underline{0.7763_{\pm 0.0090}}$ & $0.7373_{\pm 0.0020}$ & $\underline{0.7157_{\pm 0.0082}}$ \\
            AKT-Ind & $0.6362_{\pm 0.0047}$ & $0.7483_{\pm 0.0054}$ & $0.6211_{\pm 0.0046}$ & $0.6112_{\pm 0.0059}$ & $0.7702_{\pm 0.0094}$ & $0.7485_{\pm 0.0173}$ & $0.6727_{\pm 0.0191}$ & $0.6658_{\pm 0.0115}$ \\
            LPKT-Ind & $\underline{0.7344_{\pm 0.0039}}$ & $\underline{0.7579_{\pm 0.0067}}$ & $\underline{0.6541_{\pm 0.0061}}$ & $0.6296_{\pm 0.0138}$ & $0.7762_{\pm 0.0086}$ & $0.7387_{\pm 0.0179}$ & $\underline{0.7375_{\pm 0.0018}}$ & $0.7149_{\pm 0.0083}$ \\
            simpleKT-Ind & $0.6696_{\pm 0.0073}$ & $0.7481_{\pm 0.0074}$ & $0.6425_{\pm 0.0032}$ & $0.6287_{\pm 0.0076}$ & $0.7816_{\pm 0.0047}$ & $0.7566_{\pm 0.0169}$ & $0.6961_{\pm 0.0162}$ & $0.7099_{\pm 0.0127}$ \\
            \midrule
            \midrule
            RIGL & $\bm{0.7394^*_{\pm 0.0141}}$ & $\bm{0.7673^*_{\pm 0.0126}}$ & $\bm{0.7326^*_{\pm 0.0115}}$ & $\bm{0.6779^*_{\pm 0.0144}}$ & $\bm{0.8304^*_{\pm 0.0049}}$ & $\bm{0.7853^*_{\pm 0.0102}}$ &
            $\bm{0.7959^*_{\pm 0.0080}}$ & $\bm{0.7442^*_{\pm 0.0085}}$ \\
            \bottomrule
        \end{tabular}
    }
    \vspace{-4mm}
    
\end{table*}

\subsubsection{\textbf{Data Preprocessing.}} All datasets above contain group labels (i.e., the class to which the students belong), and students from the same group share the same group category. In particular, to ensure the feasibility, we conducted preprocessing on the datasets. Specifically, for each dataset, we first constructed two exercising sequences of the individual and the group based on time period divisions, where each time frame contains two types of interaction data, i.e., student-exercise responses and group-exercise responses. Particularly, for each group-exercise response under each time frame, we calculated the correct rate of this group of students on the exercise as the group's response result. We screened out groups with fewer than three students and students with less than three response logs. Due to the different temporal characteristics of the different datasets, each dataset is not divided over exactly the same time span, where ASSIST12 and NIPS-Edu are divided in days, and SLP-Math and SLP-Bio are divided in hours. Finally, since the group-exercise interactions are very sparse under each time frame in the raw data, i.e., the number of exercises answered by all students within the same group is extremely limited, we set a threshold of 0.6 to prevent the group learning sequence from being empty. 



\vspace{-2mm}
\subsection{Introduction and Implementation of Baselines}

\subsubsection{\textbf{Baselines.}} In this paper, we compare RIGL with eight baseline approaches. The details of all the comparison methods are:
\begin{itemize}[leftmargin=*]

    \item[$\bullet$] \textbf{DKT} \cite{DKT}: DKT is one of the most classical knowledge tracing methods, which utilizes a recurrent neural network (RNN) to model the exercise interaction sequences and mine the cognitive pattern between learners and exercises. 

    \item[$\bullet$] \textbf{SAKT} \cite{SAKT}: SAKT is the first knowledge tracing model introducing the self-attention mechanism~\cite{Transformer}, which exploits the transformer structure to model long-range dependencies of interaction behaviors in students' exercising sequences.

    \item[$\bullet$] \textbf{AKT} \cite{AKT}: AKT designs a novel monotonic attention module on the basis of transformer architecture for effectively modeling the forgetting behaviors of learners, which leverages an exponential decay function that can perceive contextual distance information to learn the attention weights.

    \item[$\bullet$] \textbf{LPKT} \cite{LPKT}: LPKT proposes a learning process-consistent model to explore the consistency of students' changing knowledge state during the learning process, which consists of a learning module, a forgetting module, and a predicting module.

    \item[$\bullet$] \textbf{GIKT} \cite{GIKT}: GIKT is a graph-based Interaction model for Knowledge Tracing that leverages a graph convolutional network (GCN) to effectively integrate question-skill correlations and addresses the challenge of dispersed relevant questions by considering questions and skills as different manifestations of knowledge in predicting students' mastery levels. 

    \item[$\bullet$] \textbf{simpleKT} \cite{simpleKT}: simpleKT is a simple but strong baseline method, which explicitly models the question-specific variations to capture the individual differences and uses the ordinary dot-product attention function to mine the time-aware behavior information. 

    \item[$\bullet$] \textbf{AT-DKT} \cite{AT_DKT}: AT-DKT is an novel knowledge tracing model that enhances the original deep knowledge tracing model by incorporating two auxiliary learning tasks.

    \item[$\bullet$] \textbf{DTransformer} \cite{DTransformer}: DTransformer introduces a Diagnostic Transformer with a novel contrastive learning-based training approach, designed to accurately diagnose and trace learners' knowledge proficiency.
    
\end{itemize}


\subsubsection{\textbf{Implementation.}} To adapt these baselines that only focus on individual-level knowledge tracing to the HKT task, we describe the implementation details. Specifically, for the inputs including independent learning sequence $\mathcal{R}^{\mathcal{S}}$ and the group learning sequence $\mathcal{R}^{\mathcal{O}}$, the baseline first encodes their interaction behavior, respectively, i.e., $(e^t_i,c^t_i,r^t_i)$ and $(e^t_i,c^t_i,y^t_i)$ under each time frame $\mathcal{F}_t$ and $\mathcal{H}_t$. Subsequently, the baseline performs an average aggregation of the interaction encodings under each time frame in each of the two learning sequences. Furthermore, the obtained encoding sequences are fed into the knowledge tracing module underlying the baseline for modeling the changing knowledge states of the individual and group. Finally, a joint loss function, comprising a cross-entropy loss for predicting students' performance and a mean square error loss for predicting the group's performance, is used to train the model. 

\vspace{-2mm}
\subsection{Complete Comparison Results of Individual-based Baselines}

As described in the Section~5.2, we also conducted experiments for baselines' individual version (i.e., traditional knowledge tracing model) utilizing only individual interaction data. The complete comparison results can are shown in Table~\ref{table.Result_Table}. It can be observed that in comparison to baseline models, the proposed RIGL remains significantly superior. Specifically, it shows an average increase of 5.91\% and 3.48\% over the best individual knowledge tracing baseline in terms of AUC and ACC metrics, respectively, which demonstrates the contribution of group learning behaviors to the modeling of independent learning, and the effectiveness of the proposed reciprocal learning modeling in our RIGL model.

\end{document}